# Influence of the switch-over period of an alternately active bi-heater on heat transfer enhancement inside a cavity

Anish Pal[1,#], Riddhideep Biswas[1], Sourav Sarkar[1], Aranyak Chakravarty [2,*], Achintya Mukhopadhyay[1]

[1]Department of Mechanical Engineering, Jadavpur University, Kolkata, India

[2]School of Nuclear Studies & Application, Jadavpur University, Kolkata, India

*Corresponding author: aranyak.chakravarty@jadavpuruniversity.in

#Present affiliation: Department of Mechanical Engineering, University of Illinois at Chicago, Chicago, USA

**Abstract**

Increasing power demands on multicore processors necessitate effective thermal management. The present study investigates natural convection heat transfer inside a square cavity with an alternately active bi-heater that mimics two cores of a dual-core processor. Pulsating heat flux condition is implemented on two discrete heaters with a certain switching frequency. The heat transfer characteristics have been investigated for Prandtl number =0.71 and Rayleigh number in the range of $10^3$ - $10^6$ using OpenFOAM. The results obtained for alternative active heaters configuration have been compared with that of the steady single heater and steady double-symmetric heaters subjected to the same heat flux. The alternately active heater configuration showed better heat transfer characteristics than a single steady heater for all switchover periods, and better than a double-symmetric heater for low switchover periods. However, it is found that for higher values of the switchover period, the maximum temperature of alternately active heaters configuration touches the temperature of steady single heater. This threshold switchover period has been determined using a scale analysis. The threshold switchover periods determined from scale analysis are consistent with the results obtained from numerical simulations for different Rayleigh numbers and heater lengths.

# INTRODUCTION

Technological improvements in computing systems have resulted in increasing usage of systems with multi-core processors. Along with improvement in computing performance, multi-core systems generate significantly larger amounts of heat which needs to be continuously removed. This is only possible with the help of an effective thermal management system which also ensures longevity of the system. One of the most effective means of ensuring proper thermal management of such systems is through natural convection. Other means such as liquid cooling and air-cooled heat sinks, although feasible, increase the cost as well as the system weight which hinders optimisation [1]. On the other hand, the efficiency of passive heat removal through natural convection needs to be improved in order to meet the requirements of heat removal from high power-density electronic components.

A large number of studies are available in literature which investigated natural convective heat transfer inside a rectangular cavity [2-16]. Several aspects of natural convection heat transfer, such as heater position, cavity aspect ratio, and non-uniform heat flux, have been studied in detail. However, many of these studies have considered the heaters at steady-state constant temperature. A multi-core processor, however, involves switching of jobs between different cores leading to localized pulsating heating depending on the core usage. Proper characterization of such multi-core systems, therefore, requires consideration of the transient pulsating heating of the cores. Studies on the effects of localized pulsed heating are, however, limited [17]. The resonance effect between contained natural convection and pulsating wall heating was described by Lage and Bejan [18]. Cheikh et al [19] reported the effect of aspect-ratio on natural convection in a cavity due to pulsed heating. Bae and Hyun [20] reported heat transfer enhancement due to implementation of pulsed heating in a vertical rectangular cavity with three discrete heaters. It was found that transient-stage heating temperatures could exceed corresponding steady-state values at higher Raleigh numbers. Mahapatra et al. [21] reported and quantified heat transfer enhancement associated with pulsed heating employing constant temperature conditions for the heater. Furthermore, the analysis shows that a decrease in time period results in increased heat transfer.

In the present study, a pulsating heat flux boundary condition is imposed on the heaters (instead of a constant temperature condition) inside a bottom heated square cavity. This emulates the job scheduling between the cores of a dual core processor. Study has been conducted for a range of $Ra$ ($10^4$ to $10^6$) for three different heater characteristics - alternately active heater, steady asymmetric and steady double asymmetric heater. The major objective of this work is to identify a suitable range of the switchover time period for the alternately active heaters for which the heat transfer can be augmented. Furthermore, it is worth mentioning that the limited studies that have reported heat transfer enhancement due to implementation of pulsed heating have not mentioned the minimum switch-over frequency that needs to be maintained in order to obtain the heat transfer augmentation. Proper quantification of the minimum switchover frequency is imperative because not adhering to this minimum frequency of alteration will not provide any heat-transfer augmentation. The work not only implements a pulsating heat flux boundary, which is a more authentic representation of the heaters than studies involving a temperature boundary condition, but also endeavours to enumerate the switchover frequency that keeps the maximum system temperature below the permissible limit in the cavity for any combination of $Ra$ and heater length. A rigorous scale analysis has been carried out in this work to ascertain this minimum switching frequency. This information allows the designer to determine the cooling rate of the electronic equipment with an aim to estimate the optimum switchover time for maximum possible heat transfer. The results of this analysis would provide the necessary information for better scheduling of jobs on a multi-core processor to ensure maximum heat transfer within the permissible junction temperature limit.

## PROBLEM DESCRIPTION

### *Physical configuration and assumptions*

The modelled system of heaters and the associated flow domain is depicted in Fig. 1 for the various configurations studied. The heaters represent the cores of a dual core processor and are assumed to be present on the bottom wall of a square cavity. The side walls of the cavity are kept isothermal at a lower temperature, while the top and bottom walls of the cavity are assumed to be adiabatic. All cavity walls are assumed to be rigid and impermeable.

Three different heater configurations are considered. The first configuration (Case 1, Fig. 1a) pertains to the case of alternate switching of heaters. Two heaters (H1 and H2) of equal length ($L1$ & $L2$) and placed apart at a distance $S$ ($S/H = 0.2$) are alternatively subjected to uniform heat flux i.e., at a particular instant of time, only a single heater is active. Active condition of a heater corresponds to the imposition of uniform heat flux, while in inactive state the heater is subjected to adiabatic boundary condition. The alternate activation and deactivation of the heaters is shown in Fig. 2 as a pulse graph. The second configuration (Case 2, Fig. 1b) corresponds to steady, double symmetric heaters with both heaters remaining active for the entire duration. Besides the heaters, all other conditions remain similar to Case 1. In order to ensure that equivalent thermal energy is supplied to the domain as that in Case 1, each of the two heaters are considered to be half the length ($\epsilon = 0.1$) of that considered in the Case 1 ($\epsilon = 0.2$) such that ($\sum_{i=1}^{2} L_i q_i = 0.2$). The third configuration (Case 3, Fig. 1c) considers a single steady asymmetric heater with the heater length ($\epsilon = 0.2$) being same as that in Case 1.

The working fluid is considered to be laminar and incompressible with constant isotropic and homogenous thermo-physical properties. The contributions of radiative heat transfer and viscous dissipation are neglected in the energy balance. Boussinesq approximation is utilised for modelling the natural convective effects.

*Governing Equations and Boundary conditions*

The governing equations for conservation of mass, momentum and energy in the cavity are formulated based on the assumptions made and are represented by Eqs. 1-4.

$$\frac{\partial u}{\partial x} + \frac{\partial v}{\partial y} = 0 \tag{1}$$

$$\frac{\partial u}{\partial t} + u\frac{\partial u}{\partial x} + v\frac{\partial u}{\partial y} = -\frac{1}{\rho}\frac{\partial p}{\partial x} + \frac{\mu}{\rho}\left(\frac{\partial^2 u}{\partial x^2} + \frac{\partial^2 u}{\partial y^2}\right) \tag{2}$$

$$\frac{\partial v}{\partial t} + u\frac{\partial v}{\partial x} + v\frac{\partial v}{\partial y} = -\frac{1}{\rho}\frac{\partial p}{\partial y} + \frac{\mu}{\rho}\left(\frac{\partial^2 v}{\partial x^2} + \frac{\partial^2 v}{\partial y^2}\right) - g[1 - \beta(T - T_0)] \tag{3}$$

$$\frac{\partial T}{\partial t} + u\frac{\partial T}{\partial x} + v\frac{\partial T}{\partial y} = \alpha\left(\frac{\partial^2 T}{\partial x^2} + \frac{\partial^2 T}{\partial y^2}\right) \tag{4}$$

The following scaling parameters are used to obtain the dimensionless form of the equations (Eqs. 6-9) –

$$\left.\begin{array}{l} X = \frac{x}{H}, Y = \frac{y}{H}, \quad U = \frac{uH}{\alpha}, V = \frac{vH}{\alpha}, \theta = (T - T_C)/\Delta T, \tau = \frac{t\alpha}{H^2}, Z = \frac{z\alpha}{H^2}, \Delta T = q''H/k \\ P = \frac{p^*}{\rho_0(\alpha/H)^2}, Ra = \frac{g\beta\Delta T H^3}{\nu\alpha}, Pr = \frac{\mu}{\rho C_p} \end{array}\right\} \tag{5}$$

$$\frac{\partial U}{\partial X} + \frac{\partial V}{\partial Y} = 0 \tag{6}$$

$$\frac{\partial U}{\partial \tau} + U\frac{\partial U}{\partial X} + V\frac{\partial U}{\partial Y} = -\frac{\partial P}{\partial X} + Pr\left(\frac{\partial^2 U}{\partial X^2} + \frac{\partial^2 U}{\partial Y^2}\right) \tag{7}$$

$$\frac{\partial V}{\partial \tau} + U\frac{\partial V}{\partial X} + V\frac{\partial V}{\partial Y} = -\frac{\partial P}{\partial Y} + Pr\left(\frac{\partial^2 V}{\partial X^2} + \frac{\partial^2 V}{\partial Y^2}\right) + RaPr\theta \tag{8}$$

$$\frac{\partial \theta}{\partial \tau} + U\frac{\partial \theta}{\partial X} + V\frac{\partial \theta}{\partial Y} = \left(\frac{\partial^2 \theta}{\partial X^2} + \frac{\partial^2 \theta}{\partial Y^2}\right) \tag{9}$$

Here, $X$ and $Y$ are the non-dimensional $x$ and $y$ co-ordinates. $U$ and $V$ are the non-dimensional velocity components in the $X$ and $Y$ directions, respectively. $\theta$ is the non-dimensional temperature. $\Delta T$ is the temperature scaling parameter. $\tau$ is the non-dimensional time and $P$ is the non-dimensional effective pressure. The mathematical relation between dimensionless time period and the switchover period ($z$) is also given above.

The boundary conditions for the three different heater configurations are as follows -

$$\left.\begin{aligned}&\text{Top wall:}\quad U = V = 0;\quad \partial\theta/\partial Y = 0\\&\text{Right and left wall:}\ U = V = 0; \theta = 0\\&\text{Bottom wall:}\quad U = V = 0\end{aligned}\right\} \quad (10)$$

**Case 1 (Alternative active heaters)**

$$\left.\begin{aligned}&0 < X \leq 0.2, 0.4 < X \leq 0.6, 0.8 < X \leq 1: \partial\theta/\partial Y = 0\\&0.2 < X \leq 0.4: \partial\theta/\partial Y = \begin{cases} -1, & nZ < \tau < nZ + \frac{Z}{2} \\ 0, & nZ + \frac{Z}{2} < \tau < (n+1)Z \end{cases}\\&0.6 < X \leq 0.8: \partial\theta/\partial Y = \begin{cases} -1, & nZ + \frac{Z}{2} < \tau < (n+1)Z \\ 0, & nZ < \tau < nZ + \frac{Z}{2} \end{cases}\end{aligned}\right\} \quad (11)$$

**Case 2 (Double symmetric heaters)**

$$\left.\begin{aligned}&0 < X \leq 0.3, 0.4 < X \leq 0.6, 0.7 < X \leq 1: \partial\theta/\partial Y = 0\\&0.3 < X \leq 0.4, 0.6 < X \leq 0.7: \partial\theta/\partial Y = -1\end{aligned}\right\} \quad (12)$$

**Case 3**

$$\left.\begin{aligned}&0 < X \leq 0.2, 0.4 < X \leq 1: \partial\theta/\partial Y = 0\\&0.2 < X \leq 0.4: \partial\theta/\partial Y = -1\end{aligned}\right\} \quad (13)$$

The rate of heat transfer from the heaters is quantified by the local Nusselt number at the left and right heaters, which can be defined as [8]

$$Nu = \frac{1}{\theta(X)} \qquad (14)$$

The spatial average of *Nu* on the left or right number can be defined as

$$Nu_{avg} = \frac{\int Nu \, dX}{\int dX} \qquad (15)$$

The time-average of the spatially averaged Nusselt no ($\overline{Nu}$) is accordingly defined as [21],

$$\overline{Nu} = \frac{\int_{\tau}^{\tau+Z} Nu_{avg} \, d\tau}{\int_{\tau}^{\tau+Z} d\tau} \qquad (16)$$

The time-averaged maximum non-dimensional temperature is defined as,

$$\overline{\theta_{max}} = \frac{\int_{\tau}^{\tau+Z} \theta_{max} \, d\tau}{\int_{\tau}^{\tau+Z} d\tau} \qquad (17)$$

Energy flux vectors [22] are used to visualize the transient nature of energy transport in the cavity associated with the characteristics of various heater configurations. Energy flux vectors are mathematically defined as

$$\vec{E}(X,Y) = \left(U\theta - \frac{\partial \theta}{\partial X}\right)\vec{\imath} + \left(V\theta - \frac{\partial \theta}{\partial Y}\right)\vec{\jmath} \qquad (18)$$

*Numerical Method*

The solution of the above-mentioned dimensionless governing equations (Eqs. 6-9) have been carried out using buoyantBoussinesqPimpleFoam in OpenFOAM [23]. Second-order Upwind scheme have been employed for space operators, whereas for time operator first-order schemes have been employed. A thorough grid-independent and time-independent study has been performed and a 100 X 100 grid with a time step of $10^{-5}$ has been identified to be optimum for carrying out the simulations. The code validation results corresponding to a natural convection problem have been presented in Fig. 3, where the results of present code has been validated against the results of Banerjee et al. [8].

**RESULTS AND DISCUSSION**

Heat transfer during natural convection in a square cavity with two alternately active heaters, located at the bottom of the cavity, has been studied numerically for $Pr = 0.71$ and various $Ra$ ($10^4$–$10^6$). The surface temperature and heat transfer characteristics of the alternately active heaters are compared with that of a double steady symmetric and single steady asymmetric heater at each $Ra$ for various switch over frequencies. The switchover time period ($Z$) is varied between $10^{-4}$ and 0.1.

The mechanism of fluid flow and energy transfer within the cavity for the steady heating scenarios (Case 2 and 3) can be observed from Fig. 4. In either configuration, the temperature of the working fluid rises in the vicinity of the heaters due to continuous heat exchange between the working fluid and the heater surface, while the bulk fluid temperature remains lower. This causes buoyancy-induced fluid motion within the cavity and leads to heat exchange between the heated fluid and the cold cavity walls. Two symmetric circulation cells are observed to be formed in case of the double symmetric heater configuration (Case 2). The circulation cells are observed to be asymmetric, with the larger circulation cell forming far away from the heater, for the single heater configuration (Case 3) due to asymmetric position of the heater on the bottom cavity wall.

A different mechanism is observed in case of the alternatively active heater configuration (Case 1) which can be attributed to the periodic switching of the active heaters. The transient nature of fluid flow and energy transport within the cavity for the alternatively active heaters are shown in Figs. 5 and 6 at various time instants of a complete pulsation cycle. The initial transience is neglected and $\tau = 0$ has been assigned to the instant when the initial transience is over. It can be observed from the energy flux vectors that the left heater remains active when $\tau/Z \leq 0.5$, while the right heater becomes active when $\tau/Z > 0.5$. The working fluid exchanges thermal energy with the active heater and a buoyancy-induced fluid motion develops in the cavity, similar to that observed in case of steady heating. It can be observed that when the left heater becomes active ($\tau/Z = 0.1$), the fluid circulation initially remains stronger over the left heater and the heated fluid mainly exchanges heat with the cold left side wall. As time progresses, the fluid flow is observed to bifurcate such that the heated fluid exchanges heat with both the cold sidewalls. Ultimately, the stronger fluid circulation shifts towards the right half of the cavity as the right heater becomes active ($\tau/Z = 0.6$). This mechanism repeats over time following the pulsation cycle. It can, thus, be observed that the thermal inertia of the working fluid pertaining to the previous half-cycle when the other heater was active have a significant impact on the fluid transport phenomenon associated with the half-cycle of the active heater. The energy transport mechanism for the alternatively active heater configuration, thus, differs from the steady asymmetric heater configuration, although only one heater remains active in either configuration.

*Heat Transfer Augmentation*

The effectiveness of heat transfer for the various heater configurations is compared in terms of $\overline{Nu}$. $\overline{Nu}$ for different $Z$ for each $Ra$ has been tabulated in Table 1 for the various heater configurations. An increase in $Ra$ strengthens fluid flow in the cavity leading to greater heat transfer, as indicated by the increase in $\overline{Nu}$. This is observed to be consistent for all the heater configurations. Among the heater configurations, the steady asymmetric heater (Case 3) is observed to have the most detrimental heat transfer characteristics as indicated by its lowest $\overline{Nu}$.

$\overline{Nu}$ is also observed to increase for all $Ra$ with increase in $Z$ indicating an improvement in heat transfer as switching frequency becomes larger. This results in a consequent decrease in the heater surface temperature, as shown in Fig. 7. The maximum non-dimensional surface temperature on an active heater surface is indicated by $Ra\theta_{max}$. The left heater remains active for $0 \leq \tau/Z \leq 0.5$, while the right heater is active for $0.5 \leq \tau/Z \leq 1$. This is further corroborated by the variation in time-averaged maximum surface temperature ($\overline{Ra\theta_{max}}$) as shown in Table 2.

A comparison shows that pulsed heating in the cavity results in heat transfer augmentation over a steady asymmetric heater (Case 3) for all switchover periods. Furthermore, augmentation is observed over the double steady symmetric heater configuration (Case 2) only at very low $Z$ (<=0.001). This augmentation of heat transfer in alternatively active heaters with respect to the steady asymmetric heater and for certain cases with respect to double symmetric heater is attributed to the phenomenon of periodic formation and destruction of thermal boundary layer on the heater surfaces when subjected to pulsed heating. The constant stabilization and destabilization of the thermal boundary layer can be understood by observing the transient change in isotherms in Fig. 8. The other heater follows the same characteristics. The left heater remains active for $3.41 \leq \tau \leq 3.45$ ($0 \leq \tau/Z \leq 0.5$) during which the area enclosed by a particular isotherm (for e.g. 0.04) increases with time reaching a maximum at $\tau = 3.45$ ($\tau/Z = 0.5$). This growth of the area below the isotherm indicates the formation of the thermal boundary layer near the left heater. Beyond this time, as the left heater is switched off and right heater is switched on, it is observed that the area under the isotherms continue to decrease until it ceases to exist at $\tau = 3.46$ ($\tau/Z = 0.6$) which explains the destruction of the thermal boundary layer over the left heater. A close look at the isotherms near the right heater, after it is switched on at $\tau = 3.45$ ($\tau/Z = 0.5$), also shows the simultaneous thermal boundary layer growth and destruction over the right heater. It is observed that heat transfer augmentation is obtained for the cases where the maximum thickness of the thermal boundary layer for the alternatively active heater configuration remains lower as compared to that obtained in the corresponding situations for Case 2 and 3.

*Estimating the threshold switchover frequency*

However, as observed from Fig 7, for certain switch-over periods the maximum surface temperature ($Ra\theta_{max}$) reaches the surface temperature of a steady asymmetric heater. This is a significant material constraint. Thus, although implementation of any pulsation frequency augments heat transfer, not all frequencies ensure that the maximum heater temperature will remain within this allowable limit. If the time period for formation of fully developed boundary layer ($\tau_f$) over the active heater during the heating part of the cycle is of the same order as of half-switch over period ($Z/2$), the thermal boundary layer over the active heater will achieve fully developed state within the heater activation time period. This allows the active heater maximum temperature to reach that of steady asymmetric heater. Hence, a switchover period shorter than the limiting time period (i.e., $Z < 2\tau_f$) ensures that the heater temperature remains within the allowable limit.

A scale analysis has been carried out to estimate $\tau_f$ and hence, determine the threshold switchover period ($Z_{th} = 2\tau_f$). It is assumed that a single switching cycle for a single active heater can be subdivided into two periods – an initial transient period where conduction heat transfer dominates and a steady heating period when convection heat transfer becomes important. This demarcation is determined to be the time period taken for the thermal boundary layer to become fully developed. In the initial transient period, the fluid near the active heater remains stationary immediately after the heater becomes active and the entire energy transfer to the fluid is through conduction. Thus, neglecting convective heat transfer in Eq. 4, thermal inertia scales with thermal diffusion such that

$$\frac{\Delta T}{t} \sim \alpha \frac{\Delta T}{L_c \delta_T} \Rightarrow \delta_T \sim \frac{\alpha t}{L_c} \qquad (19)$$

The velocity scale during this initial transient period can be obtained by eliminating the pressure terms in Eq. 2 and 3. Taking derivative of Eq. 2 with respect to $y$ and of Eq. 3 with respect to $x$, and subtracting the resulting derivatives, we can eliminate the pressure terms as shown in Eq. 20.

$$\frac{\partial}{\partial x}\left(\frac{\partial v}{\partial t} + u\frac{\partial v}{\partial x} + v\frac{\partial v}{\partial y}\right) - \frac{\partial}{\partial y}\left(\frac{\partial u}{\partial t} + u\frac{\partial u}{\partial x} + v\frac{\partial u}{\partial y}\right)$$

$$= \frac{\mu}{\rho}\left[\frac{\partial}{\partial x}\left(\frac{\partial^2 v}{\partial x^2} + \frac{\partial^2 v}{\partial y^2}\right) - \frac{\partial}{\partial y}\left(\frac{\partial^2 u}{\partial x^2} + \frac{\partial^2 u}{\partial y^2}\right)\right] + g\beta\frac{\partial T}{\partial x} \qquad (20)$$

Equation 20 has three primary groups of terms: inertia terms on the left-hand side, and viscous diffusion and buoyancy terms on the right-hand side. Since $x \sim H$, $y \sim \delta_T$ and $\delta_T \ll H$, all x-derivate terms are neglected, leading to the following terms dominating each group –

*Inertia*   *Friction*   *Buoyancy*

$$\frac{\partial^2 u}{\partial y \, \partial t}, \qquad \frac{\mu}{\rho}\frac{\partial^3 u}{\partial y^3}, \qquad g\beta\frac{\partial T}{\partial x} \qquad (21)$$

Scaling with respect to the friction term -

$$\frac{\tau}{Pr}, \qquad 1, \qquad \frac{g\beta\Delta T\delta_T^3}{\nu u H} \qquad (22)$$

For fluids with $Pr \geq 1$ and since $\tau < 1$ in the initial transient period, at t > 0 buoyancy is balanced by fluid friction. This allows us to obtain an initial horizontal velocity scale as

$$u \sim \frac{g\beta\Delta T\alpha^3 t^3}{\frac{\mu}{\rho}\epsilon^3 H^4} \qquad (23)$$

As time increases, the impact of convective heat transfer becomes larger, whereas the influence of inertia reduces in comparison. This transience continues till the thermal boundary layer becomes fully developed, after which there exists a balance between the heat transmitted from the wall and the enthalpy transported away by the buoyant fluid layer. In this situation, heat transfer through convection scales as the conduction heat transfer as

$$u\frac{\Delta T}{H} \sim \alpha \frac{\Delta T}{\delta_T^2} \Longrightarrow u \sim \alpha \frac{H}{\delta_T^2} \qquad (24)$$

Thus, we obtain two velocity scales in the transient and steady periods. At t = $t_f$, the transient velocity scale should therefore, scale as the steady velocity scale i.e.

$$\frac{g\beta \Delta T \alpha^3 t_f^3}{\frac{\mu}{\rho}\epsilon^3 H^4} \sim \alpha \frac{H}{\delta_T^2} \qquad (25)$$

This can be re-written in terms of $t_f$ as

$$t_f \sim \epsilon \frac{H^2}{\alpha} Ra^{-\frac{1}{5}} \qquad (26)$$

In dimensionless form, this can be written as

$$\tau_f \sim \epsilon Ra^{-\frac{1}{5}} \qquad (27)$$

The obtained scaled $Z_{th}(= 2\tau_f)$ which is dependent on both the energy supplied ($Ra$) and the heater length ($\epsilon$) serves as the threshold switch-over period for ensuring that the material constraint temperature is not breached. Numerically, $Z_{th}$ is determined from the variation of $Ra\theta_{max}$ with $Z$, as shown for some representative cases in Fig. 9. The authenticity of equation 18 for various combinations of Ra and $\epsilon$ is depicted through the plot of $\text{Log}_{10}\left(\frac{Z_{th}}{\epsilon}\right)$ vs $\text{Log}_{10}$ (Ra) in Fig. 10 . All the numerically obtained data points pertaining to various combinations of Ra and $\epsilon$ for the above mentioned Log-Log plot collapse on a line with slope of $-\frac{1}{5}$ thus proving the validity of the scaling law $\tau_f \sim \epsilon Ra^{-\frac{1}{5}}$ established earlier. Hence, it can be concluded that $Z < 2\epsilon Ra^{-\frac{1}{5}}$ must be maintained for ensuring that the maximum active heater surface temperature remains below than that of steady active heater. A detailed numerical investigation reveals that Z less than that of $2\xi\epsilon Ra^{-\frac{1}{5}}$, $\xi$ being the scale factor with magnitude of 2.43 needs to be maintained for ensuring that the maximum surface temperature constraint is not breached

## CONCLUSIONS

Natural convection in an cavity corresponding to single steady asymmetric heater, double symmetric heater and alternately active heater considering heat flux as the boundary condition with various switchover time periods (0.1 to 10$^{-4}$) for $Ra$ in the range of 10$^3$ to 10$^6$ and $Pr$ = 0.71 has been investigated. The impacts of switching time period on surface temperature and heat transport are investigated for alternately active heaters.

It has been found in the present work that the heat transfer enhances with the decrease of switchover time period ($Z$) and has been found to be higher than that of the steady state value in all cases for all $Ra$. Periodic formation and destruction of thermal boundary layer leads to this heat transfer augmentation. Just as heat transfer improves with the increase in switching frequency, the surface temperature of the active heater remains below that of asymmetric steady state heater for any $Ra$. A maximum increase of approximately 66.66% and 43% of the temporally averaged $Nu$ and maximum surface temperature over the steady asymmetric heater has been observed for Z = 0.0001.

However, one interesting factor found in this investigation is that for certain switchover periods the maximum surface temperature of the active pulsating heater reaches that of the surface temperature of the steady asymmetric heater and this is a significant material constraint. If the half-switchover period is of the same order as of the time required for the formation of fully developed thermal boundary layer, the maximum surface temperature of the active heater reaches that of the steady heater. This phenomenon of the maximum surface temperature of the active heater reaching that of the steady-heater in spite of implementation of pulsed heating has been observed numerically for various combinations of $Ra$ and $\epsilon$. Hence, the switch-over time period must be maintained properly. Scale analysis yields the time of formation of fully developed thermal boundary layer, $\tau_f \sim \epsilon Ra^{-1/5}$. A detailed numerical investigation reports that the above discussed threshold switch over period needs to be maintained ξ times that of scaled $Z_{th}$. ξ, being the scale-factor having magnitude 2.43.

Improvement of heat transfer characteristics with $Ra$ for any time period gives an indication that judicial choice of job switching frequency can cater to the need of effective thermal management of ever increasing processor power. Furthermore maintaining a minimum switching frequency ensures that the maximum heater surface temperature remains within an allowable limit.

Table 1: Variation of $\overline{Nu}$ on the active heater for various $Ra$.

| $Ra$ \ $Z$ | Case 1 | | | | | Case2 | Case3 |
|---|---|---|---|---|---|---|---|
| | 0.1 | 0.05 | 0.01 | 0.001 | 0.0001 | | |
| $10^4$ | 7.29 | 7.89 | 9.089 | 10.01 | 10.32 | 6.51 | 9.38 |
| $10^5$ | 8.55 | 9.24 | 10.89 | 12.26 | 12.72 | 7.84 | 10.65 |
| $10^6$ | 12.41 | 12.99 | 16.16 | 19.4 | 20.53 | 12.28 | 15.87 |

Table 2: Variation of $Ra\overline{\theta_{max}}$ for different $Ra$ on the active heater

| $Z$ \ $Ra$ | | $10^4$ | $10^5$ | $10^6$ |
|---|---|---|---|---|
| Case 1 | 0.1 | 1.52 x $10^3$ | 1.34 x $10^4$ | $10^5$ |
| | 0.05 | 1.41 x $10^3$ | 1.24 x $10^4$ | 9.5 x $10^4$ |
| | 0.01 | 1.21 x $10^3$ | 1.05 x $10^4$ | 7.4 x $10^4$ |
| | 0.001 | 1.09 x $10^3$ | 0.93 x $10^4$ | 6.2 x $10^4$ |
| | 0.0001 | 1.06 x $10^3$ | 0.9 x $10^4$ | 5.9 x $10^4$ |
| Case 2 | | 1.13 x $10^3$ | 1.013 x $10^4$ | 7.1 x $10^4$ |
| Case 3 | | 1.678 x $10^3$ | 1.44 x $10^4$ | 1.016 x $10^5$ |

# List of Figures

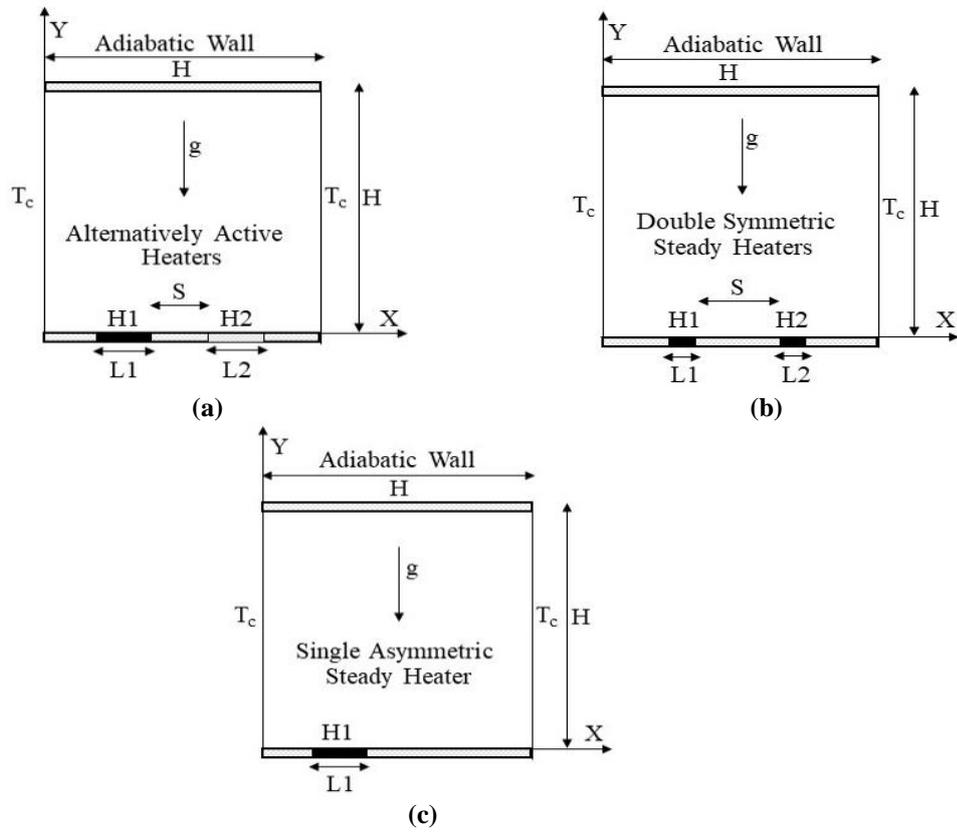

**Figure 1: Schematic showing computational domain for (a) alternatively active heaters (Case1), (b) double symmetric steady heaters (Case 2) and (c) single asymmetric steady heater (Case 3).**

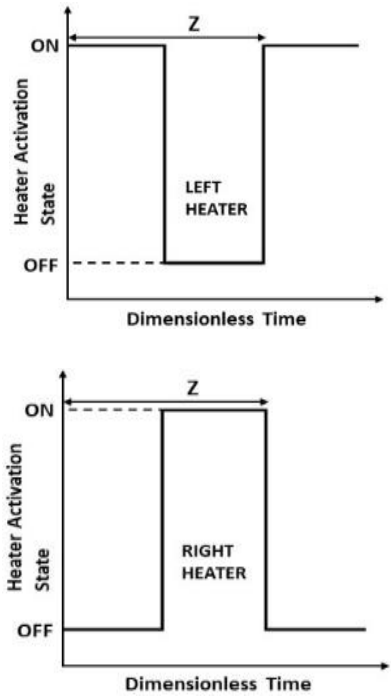

**Figure 2: Pulse graph of Left (top) and Right Heater (bottom)**

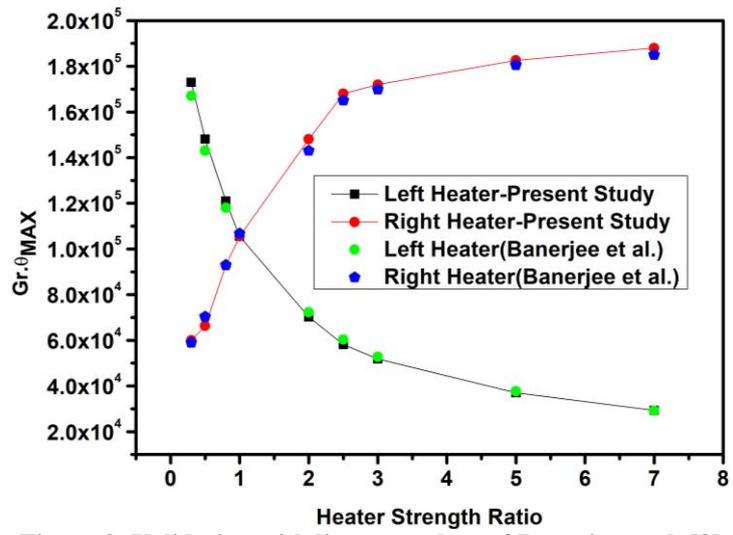
**Figure 3: Validation with literature data of Banerjee et al. [8]**

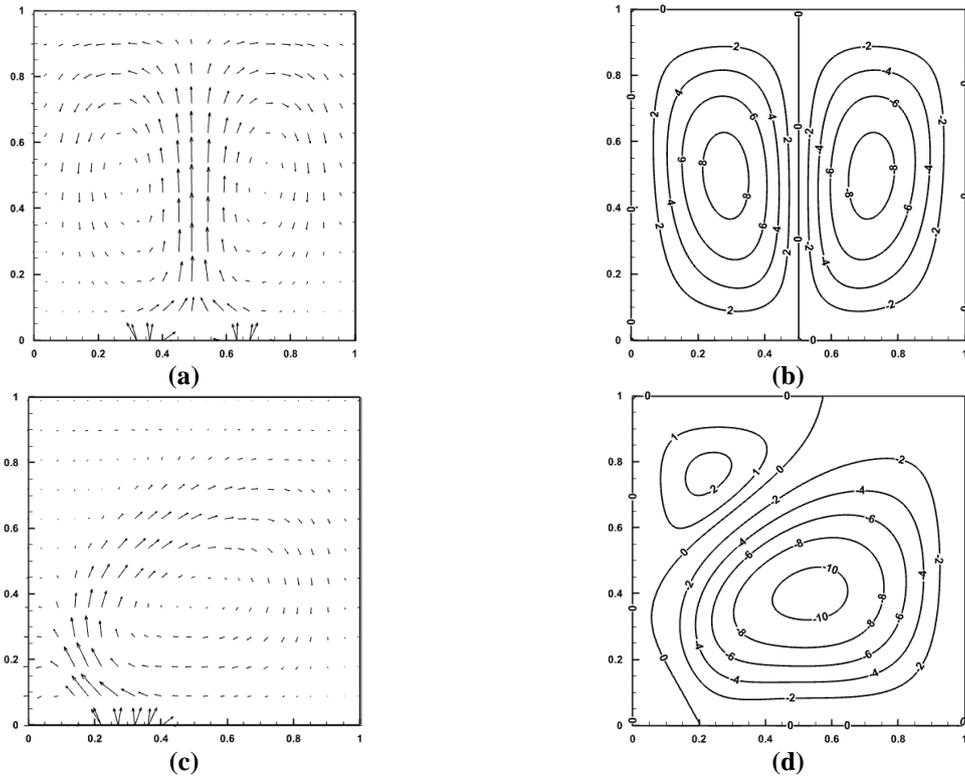

**Figure 4 (a)-(b): Energy vectors (right) and streamlines (left) for double symmetric heater (case 2)**
**(c)-(d): Energy vectors (right) and streamlines (left) for single asymmetric heater ( case 3)**

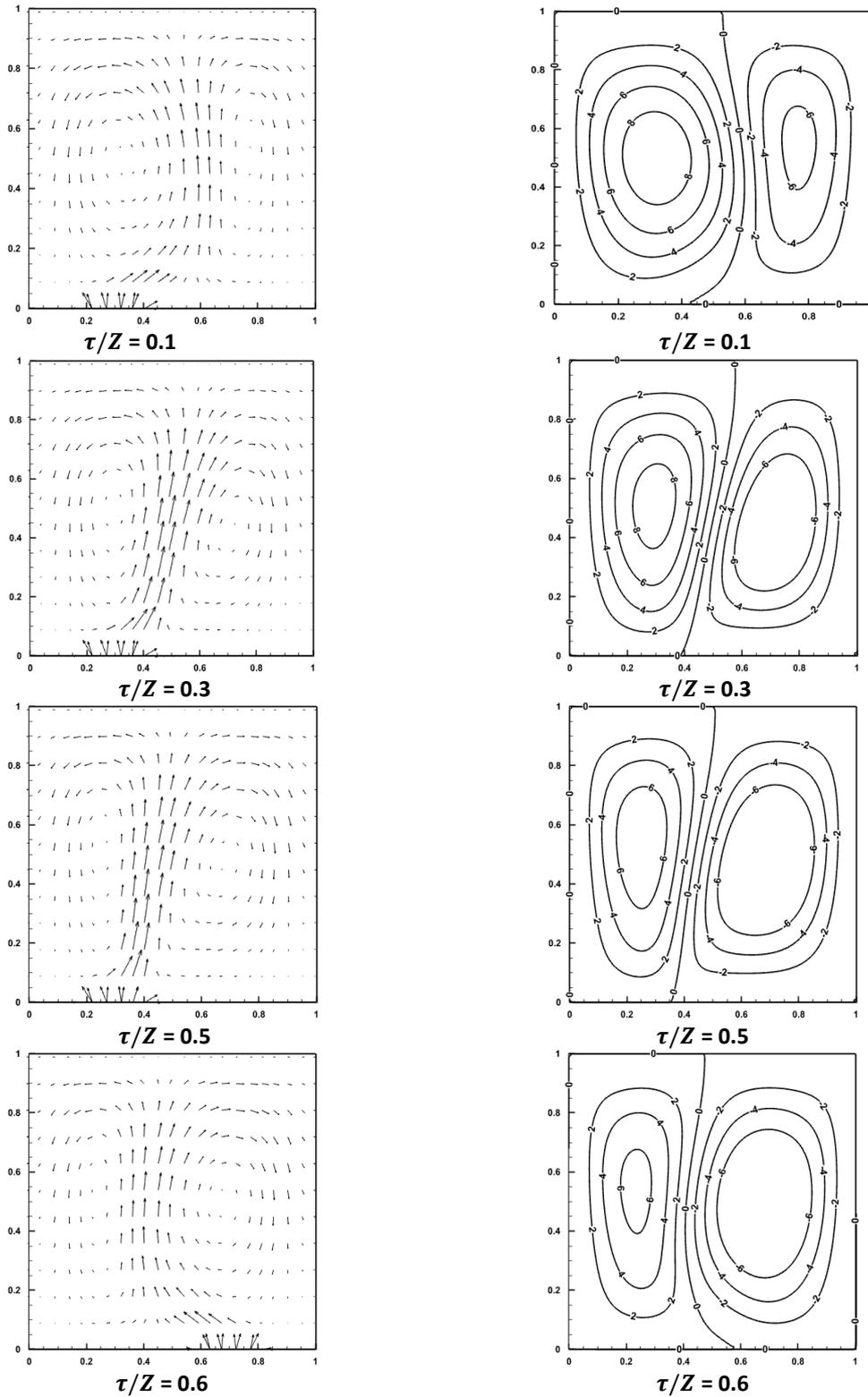

**Figure 5:** Energy flux vectors (right) and streamlines (left) at various time instants of a cycle ( up to $\tau/Z$ = 0.6)( Case 1)

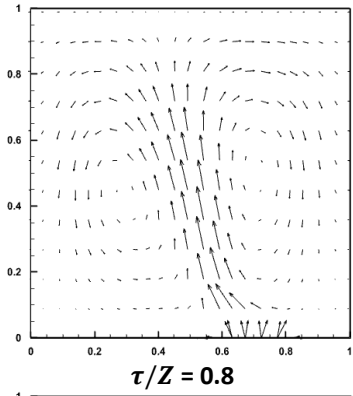 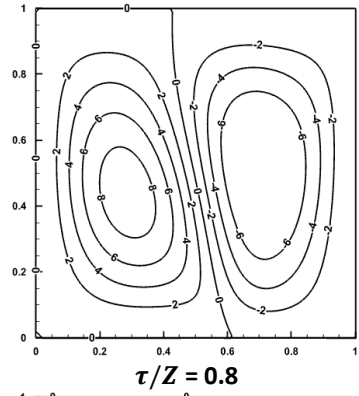
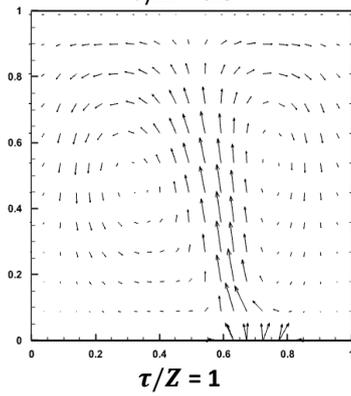 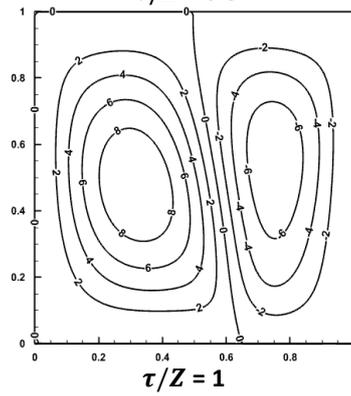

**Figure 6: Energy vectors (right) and streamlines (left) at various time instants of a cycle ( from $\tau/Z$ = 0.8-1.0)( Case 1)**

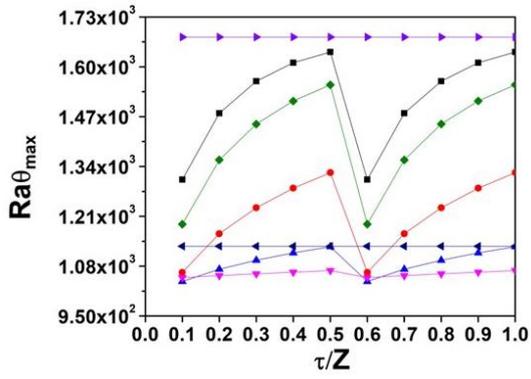
(a) $Ra = 10^4$

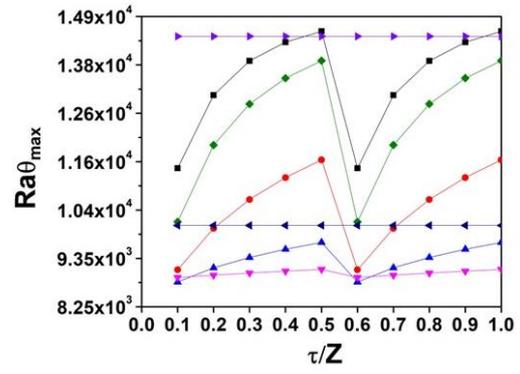
(b) $Ra = 10^5$

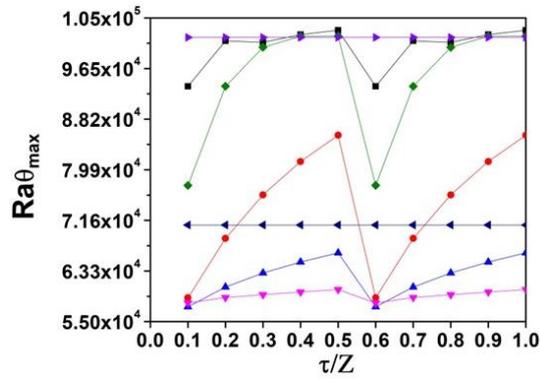
(c) $Ra = 10^6$

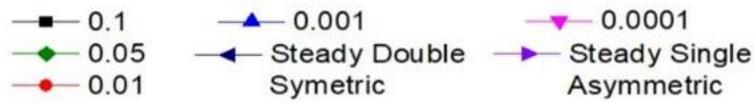

**Figure 7 (a)-(c): Variation of $Ra\theta_{max}$ of the active heater with $Z$ for different $Ra$.**

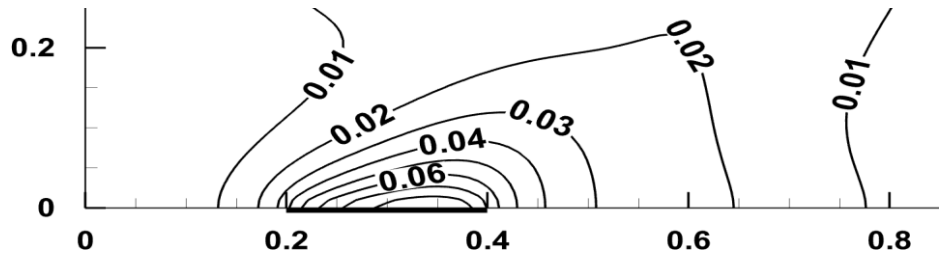

**(a)** $\tau = 3.41$

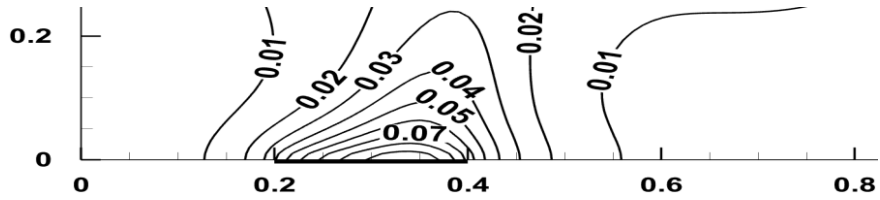

**(b)** $\tau = 3.45$

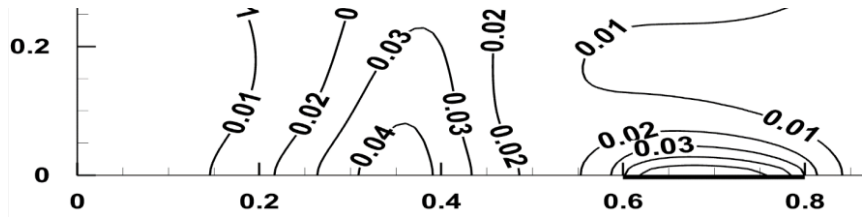

**(c)** $\tau = 3.453$

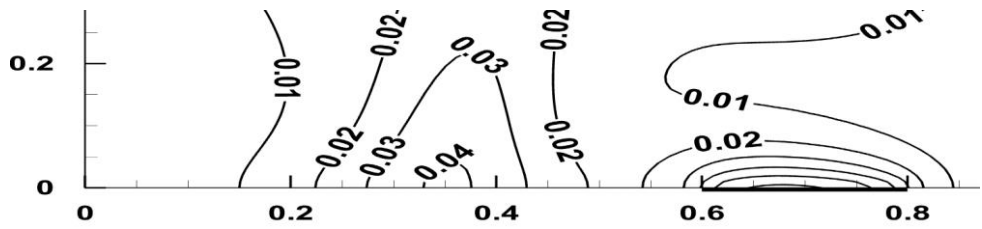

**(d)** $\tau = 3.454$

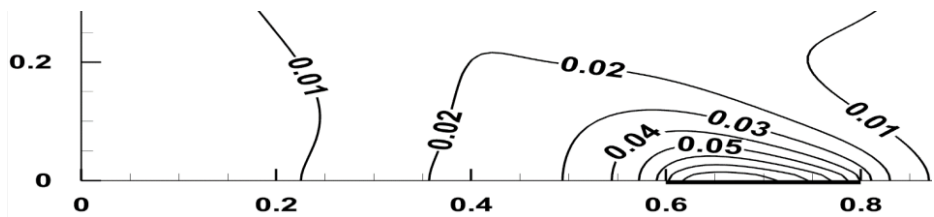

**(e)** $\tau = 3.46$

**Figure 8: Isotherms at different time instants for $Z = 0.1$, $Ra = 10^6$.**

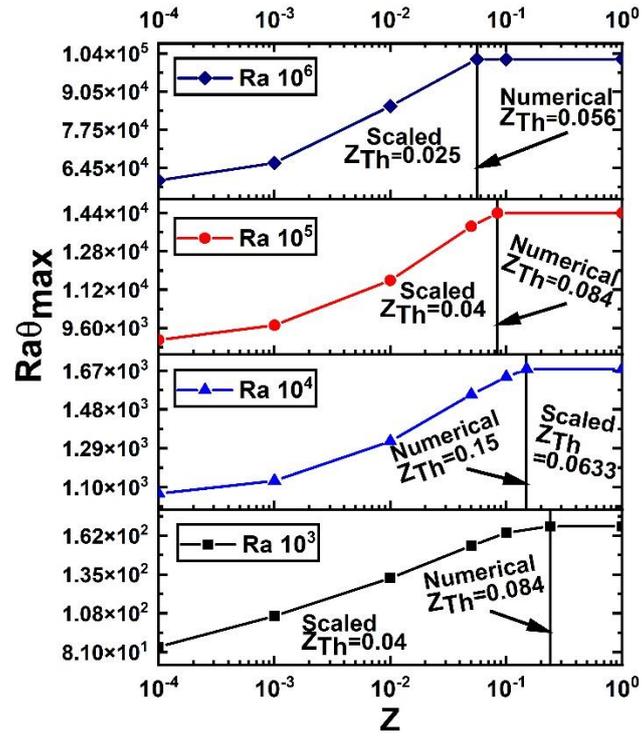

**Figure 9: Variation of active heater maximum temperature with switchover period of the alternatively active heaters ($\epsilon = 0.2$).**

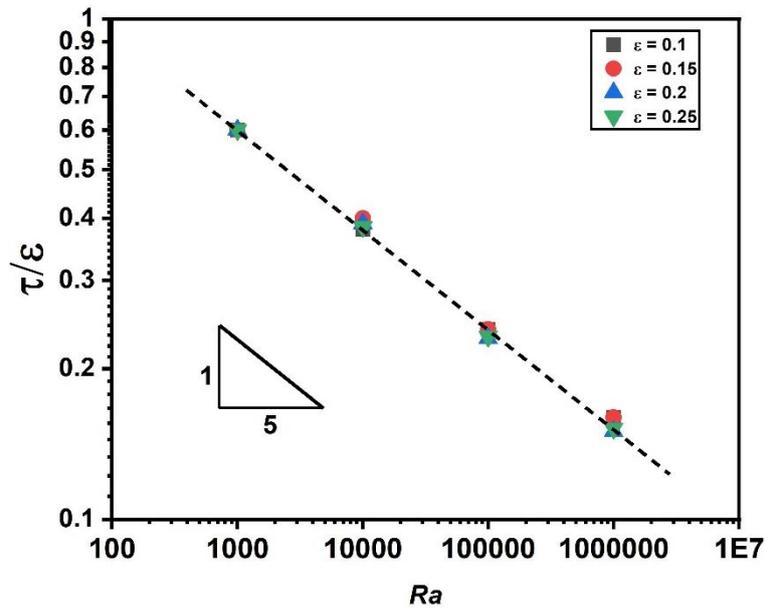

**Figure 10:** Variation of the ratio of the numerically obtained $Z_{th}$ with the scaled $Z_{th}$ for various $Ra$ and $\epsilon$.